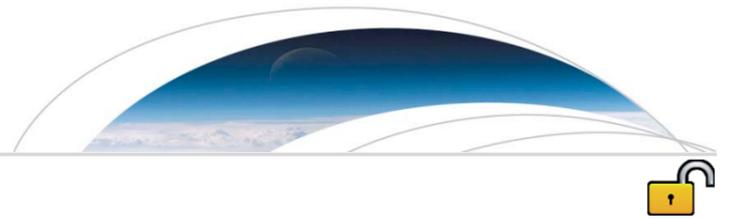

# Birkeland current effects on high-latitude ground magnetic field perturbations

K. M. Laundal[1,2], S. E. Haaland[1,3], N. Lehtinen[1], J. W. Gjerloev[1,4], N. Østgaard[1], P. Tenfjord[1], J. P. Reistad[1], K. Snekvik[1], S. E. Milan[1,5], S. Ohtani[4], and B. J. Anderson[4]

[1]Birkeland Centre for Space Science, University of Bergen, Bergen, Norway, [2]Teknova AS, Kristiansand, Norway, [3]Max Planck Institute for Solar System Research, Göttingen, Germany, [4]The Johns Hopkins University Applied Physics Laboratory, Laurel, Maryland, USA, [5]Department of Physics and Astronomy, University of Leicester, Leicester, UK

**Abstract** Magnetic perturbations on ground at high latitudes are directly associated only with the divergence-free component of the height-integrated horizontal ionospheric current, $J_{\perp,df}$. Here we show how $J_{\perp,df}$ can be expressed as the total horizontal current $J_{\perp}$ minus its curl-free component, the latter being completely determined by the global Birkeland current pattern. Thus, in regions where $J_{\perp} = 0$, the global Birkeland current distribution alone determines the local magnetic perturbation. We show with observations from ground and space that in the polar cap, the ground magnetic field perturbations tend to align with the Birkeland current contribution in darkness but not in sunlight. We also show that in sunlight, the magnetic perturbations are typically such that the equivalent overhead current is antiparallel to the convection, indicating that the Hall current system dominates. Thus, the ground magnetic field in the polar cap relates to different current systems in sunlight and in darkness.

## 1. Introduction

The magnetic field perturbations on ground $B_{gnd}$ are associated with a divergence-free current system [e.g., *Vasyliunas*, 1999], often represented as a two-dimensional horizontal "equivalent" current. Several equivalent current systems exist [e.g., *Chapman and Bartels*, 1940; *Kamide*, 1982; *Amm*, 1997], the most simple being a rotation of the horizontal magnetic field perturbation to align with an overhead line current, $J_{eq} \propto B_{gnd} \times \hat{z}$ where $\hat{z}$ is an upward unit vector. It is important to note that the equivalent currents are mathematical constructions which are not necessarily equal to the actual currents. It is the aim of this paper to show how various physical current systems driven by the solar-wind-magnetosphere interaction relate to the equivalent currents at high latitudes and thus the magnetic disturbance field on ground.

The ionospheric current system can be decomposed into Birkeland currents ($J_{\parallel}$) which are parallel with the magnetic field and height-integrated horizontal currents ($J_{\perp}$). The latter can be further decomposed into divergence-free and curl-free (potential) components, $J_{\perp} = J_{\perp,df} + J_{\perp,cf}$ (Helmholtz decomposition). In regions where the magnetic field lines are radial, a Biot-Savart integral over such a current system will yield zero contribution from all but the horizontal divergence-free component [*Vasyliunas*, 2007], a result which is known as Fukushima's theorem [e.g., *Fukushima*, 1994]. This property holds to a good approximation as long as $\sin \chi$ is close to unity, $\chi$ being the inclination of the Earth's magnetic field [*Untiedt and Baumjohann*, 1993]. In the polar cap, which is the main area of interest in the present paper, Fukushima's theorem can be safely applied, since at >70°, $\sin \chi > 0.98$ in a dipole field. Note that the discussion above is independent of conductivity. Often, when Fukushima's theorem is introduced, it is applied to Hall and Pedersen currents, in which case the conductivity must be considered as well. This will be discussed further in the next section.

Consequently, we can set $J_{eq} = J_{\perp,df}$. In the following we show, for a flat geometry, how Birkeland currents may influence the decomposition into divergence-free and curl-free components and thus also the ground magnetic field perturbations. We also discuss the influence of Hall and Pedersen currents on $J_{\perp,df}$. In section 3 we compare ground-based magnetic field measurements from SuperMAG [*Gjerloev*, 2012] with space-based magnetic field measurements and associated Birkeland currents from the Active Magnetosphere and Planetary Electrodynamics Response Experiment (AMPERE) [*Anderson et al.*, 2000; *Waters et al.*, 2001] and electric field measurements from Cluster Electron Drift Instrument (EDI) [*Paschmann et al.*, 1997], to test the results observationally.

## 2. Birkeland Current Influence on the Divergence-Free Currents

In this section we derive an expression for the divergence-free component of the horizontal current in terms of the total horizontal current and Birkeland currents. We will show that the Birkeland currents indirectly affect the equivalent current, and thus ground magnetic perturbations, by affecting how $\mathbf{J}_\perp$ is decomposed into $\mathbf{J}_{\perp,cf}$ and $\mathbf{J}_{\perp,df}$.

In a flat geometry we place $\mathbf{J}_\perp$ at $z = 0$ and $J_\parallel$ along $\hat{\mathbf{z}}$ at $z > 0$:

$$\mathbf{J} = \delta(z)\mathbf{J}_\perp + \hat{\mathbf{z}}H(z)J_\parallel = \delta(z)(\nabla_\perp \alpha + \mathbf{J}_{\perp,df}) - \hat{\mathbf{z}}H(z)\nabla_\perp^2 \alpha \quad (1)$$

where $\delta(z)$ is Dirac's delta function. $H(z) = 1$ if $z > 0$, and 0 otherwise. The curl-free component of the horizontal current system is written on the right hand side as the gradient of a scalar potential, $\mathbf{J}_{\perp,cf} = \nabla_\perp \alpha$, and we have implied current continuity:

$$\nabla \cdot \mathbf{J}_\perp = \nabla_\perp^2 \alpha = -J_\parallel. \quad (2)$$

Equation (2) can be solved for $\alpha$ using the Green function for the 2-D Laplacian:

$$\alpha(\mathbf{r}_\perp) = \int d\mathbf{r}'_\perp G(r_\perp - r'_\perp)\nabla'_\perp \cdot \mathbf{J}_\perp(\mathbf{r}'_\perp)$$
$$= -\int d\mathbf{r}'_\perp G(r_\perp - r'_\perp)J_\parallel(\mathbf{r}'_\perp) \quad (3)$$

where the integral is over $\mathbf{r}'_\perp$ and evaluated at $\mathbf{r}_\perp$ (both are vectors in the $z = 0$ plane), and

$$G(r_\perp - r'_\perp) = \frac{1}{2\pi} \ln |r_\perp - r'_\perp|. \quad (4)$$

Thus, we get the following expression for the equivalent current $\mathbf{J}_{eq}$ in terms of the horizontal current $\mathbf{J}_\perp$ and the Birkeland currents $J_\parallel$:

$$\mathbf{J}_{eq} = \mathbf{J}_{\perp,df} = \mathbf{J}_\perp - \mathbf{J}_{\perp,cf} = \mathbf{J}_\perp - \nabla_\perp \alpha \quad (5)$$

where

$$-\nabla_\perp \alpha(\mathbf{r}_\perp) = \frac{1}{2\pi}\nabla_\perp \int d\mathbf{r}'_\perp \ln |r_\perp - r'_\perp| J_\parallel(\mathbf{r}'_\perp), \quad (6)$$

which means that if we know the divergence of the horizontal currents (i.e., $-J_\parallel$) in addition to $\mathbf{J}_{\perp,df}$, the full current system can be found.

This equation also shows that in regions with zero horizontal currents (and consequently zero $J_\parallel$), which may be a reasonable assumption in the polar cap when it is dark, the equivalent current is completely determined by the global pattern of Birkeland currents. Then we have that $\mathbf{J}_{eq} = \mathbf{J}_{\perp,df} = -\mathbf{J}_{\perp,cf} = -\nabla_\perp \alpha$. From equation (6) we see that the direction of $-\mathbf{J}_{cf}$ is downward (negative) Birkeland currents and away from upward (positive) Birkeland currents. For the typical Region 1 current system, it points dawnward in the polar cap.

The fact that the equivalent current and thus the corresponding magnetic field on ground, $\mathbf{B}_{gnd}$, in some cases can be expressed as a function of $J_\parallel$ seems to violate Fukushima's theorem. However, this is not the case since the magnetic effect of $\nabla_\perp \alpha$ is zero, as can be seen by writing the Biot-Savart law on the following form:

$$\mathbf{B}_{gnd} = -\frac{\mu_0}{4\pi} \int d^3r' \frac{\nabla' \times \mathbf{J}(\mathbf{r}')}{|\mathbf{r} - \mathbf{r}'|}, \quad (7)$$

which shows that $\mathbf{B}_{gnd}$ is independent of currents whose curl is zero, which is true for $\nabla_\perp \alpha$. We end up with the apparent conundrum that the equivalent currents deduced from the magnetic disturbance field depend on $J_\parallel$ (equations (5) and (6)), but the magnetic disturbance field itself does not (equation (7)). Notice, however, that the integral in equation (7) will have zero contribution from regions where $\mathbf{J}_\perp = 0$, independent of $\nabla_\perp \alpha$, but the value of $\nabla_\perp \alpha$ in these regions are not necessarily zero, since it depends on the global divergence-free current. $\nabla_\perp \alpha$ is what *Untiedt and Baumjohann* [1993] termed "fictitious horizontal closure currents."

These results show that $J_\parallel$ indirectly affects $\Delta \mathbf{B}$ on ground, by changing how $\mathbf{J}_\perp$ is decomposed into $\mathbf{J}_{\perp,cf}$ and $\mathbf{J}_{\perp,df}$, the latter being observable from ground.

### 2.1. Relation to Hall and Pedersen Currents

It is common to express the height-integrated horizontal currents as the sum of Hall and Pedersen currents, $\mathbf{J}_\perp = \mathbf{J}_H + \mathbf{J}_P$, which are defined relative to the electric field: Pedersen currents are parallel to the electric field, and Hall currents are parallel to $\mathbf{B} \times \mathbf{E}$. This decomposition is physically more meaningful compared to $\mathbf{J}_{\perp,df}$ and $\mathbf{J}_{\perp,cf}$, since (1) the components refer to the electric field which in turn is produced by the magnetosphere-ionosphere plasma dynamics [e.g., *Parker*, 1996] and (2) $\mathbf{J}_H$ and $\mathbf{J}_P$ are everywhere orthogonal and thus components of the actual current. That implies that if $\mathbf{J}_\perp = 0$, $\mathbf{J}_H = \mathbf{J}_P = 0$, in contrast to $\mathbf{J}_{\perp,df}$ and $\mathbf{J}_{\perp,cf}$ which may be nonzero but whose sum must cancel. However, the relation of $\mathbf{J}_H$ and $\mathbf{J}_P$ to ground magnetic field perturbations is more complicated than for $\mathbf{J}_{\perp,df}$ and $\mathbf{J}_{\perp,cf}$.

In special cases, the Hall current coincides with the divergence-free current and the Pedersen current with the curl-free current. Using Ohm's law, and writing the electric field as a potential $\mathbf{E} = -\nabla\Phi$, they are

$$\mathbf{J}_P = -\Sigma_P \nabla\Phi \tag{8}$$

$$\mathbf{J}_H = -\Sigma_H \nabla\Phi \times \hat{\mathbf{z}}, \tag{9}$$

where $\Sigma_P$ and $\Sigma_H$ are Pedersen and Hall conductances (height-integrated conductivities), respectively. In the Southern Hemisphere, the sign of equation (9) is reversed. To show when $\mathbf{J}_H = \mathbf{J}_{\perp,df}$ and $\mathbf{J}_P = \mathbf{J}_{\perp,cf}$, we calculate the divergence:

$$\nabla \cdot \mathbf{J}_H = \hat{\mathbf{z}} \cdot (\nabla\Sigma_H \times \nabla\Phi) \tag{10}$$

$$\nabla \cdot \mathbf{J}_P = -\Sigma_P \nabla^2\Phi - \nabla\Phi \cdot \nabla\Sigma_P \tag{11}$$

and the curl:

$$\nabla \times \mathbf{J}_H = \Sigma_H \nabla^2\Phi \hat{\mathbf{z}} - \nabla\Sigma_H (\nabla\Phi \times \hat{\mathbf{z}}) \tag{12}$$

$$\nabla \times \mathbf{J}_P = -\nabla\Sigma_P \times \nabla\Phi \tag{13}$$

From these equations we see that the Pedersen current is curl-free and the Hall current divergence-free if $\nabla\Sigma_H \cdot \mathbf{v} = \nabla\Sigma_P \cdot \mathbf{v} = 0$ where $\mathbf{v}$ is the convection velocity, which is parallel to contours of constant $\Phi$. If these conditions are fulfilled everywhere, $\mathbf{J}_P = \mathbf{J}_{\perp,cf}$ and $\mathbf{J}_H = \mathbf{J}_{\perp,df} = \mathbf{J}_{eq}$, since the Helmholtz decomposition is unique, assuming $\mathbf{J}_\perp$, $\nabla \cdot \mathbf{J}_\perp$ and $\nabla \times \mathbf{J}_\perp$ go to zero at low latitudes. Note that this will necessarily be true if the conductances are uniform, since their gradients are zero then. It can, however, be true even with nonzero conductance gradients, as long as the gradient is perpendicular to equipotential contours. This is typically the case on the dawn and dusk polar cap boundaries, since plasma flow only crosses this boundary in the reconnection regions [e.g., *Cowley and Lockwood*, 1992], and the conductance gradients point from the polar cap to the auroral zone where ionization from particle precipitation is prevalent.

There may be cases when $\mathbf{J}_H$ and $\mathbf{J}_P$ are divergence- and curl-free, respectively, only in localized regions. Then we have, in general, that $\mathbf{J}_H \neq \mathbf{J}_{\perp,df}$ and $\mathbf{J}_P \neq \mathbf{J}_{\perp,cf}$ even in regions where $\nabla \times \mathbf{J}_P = 0$ and $\nabla \cdot \mathbf{J}_H = 0$. For example, in regions where $\Sigma_P \approx \Sigma_H \approx 0$, $\mathbf{J}_\perp = \mathbf{J}_H = \mathbf{J}_P = 0$, so that $\mathbf{J}_H$ necessarily is divergence-free and $\mathbf{J}_P$ curl-free. $\mathbf{J}_{\perp,df}$ will be equal to $-\mathbf{J}_{\perp,cf}$ in such regions, but they are not necessarily zero.

Thus the interpretation of equivalent currents is highly dependent on the conductance distribution. The following summarize two extremes:

Case 1: $\Sigma_H = \Sigma_P = 0$ (locally): $\mathbf{J}_\perp = 0$, and $\mathbf{J}_{eq} = -\nabla_\perp \alpha$, which is completely determined by the global Birkeland currents. If this holds in the polar cap, the equivalent current will be dawnward for a typical R1 current system, and the ground magnetic field perturbations will be sunward. The magnetic field perturbations will thus point roughly in the same direction in space and on ground.

Case 2: $\nabla\Sigma_H \cdot \mathbf{v} = \nabla\Sigma_P \cdot \mathbf{v} = 0$ (globally) and $\Sigma_H, \Sigma_P$ nonzero: $\mathbf{J}_P = \nabla_\perp \alpha$ and $\mathbf{J}_H = \mathbf{J}_{eq}$. The equivalent currents flow antiparallel to the ionospheric convection ($\mathbf{v}$), and perpendicular to $-\nabla_\perp \alpha$ which exactly balances the Pedersen current. The magnetic field perturbations on ground will be perpendicular to magnetically conjugate perturbations in space.

The former prediction is expected to largely apply in the polar cap when it is dark. Then the conductivity is close to zero, and the horizontal currents weak [*Gjerloev and Hoffman*, 2014]. In sunlight, the situation is expected to be closer to the latter situation. We test these predictions in the next section.

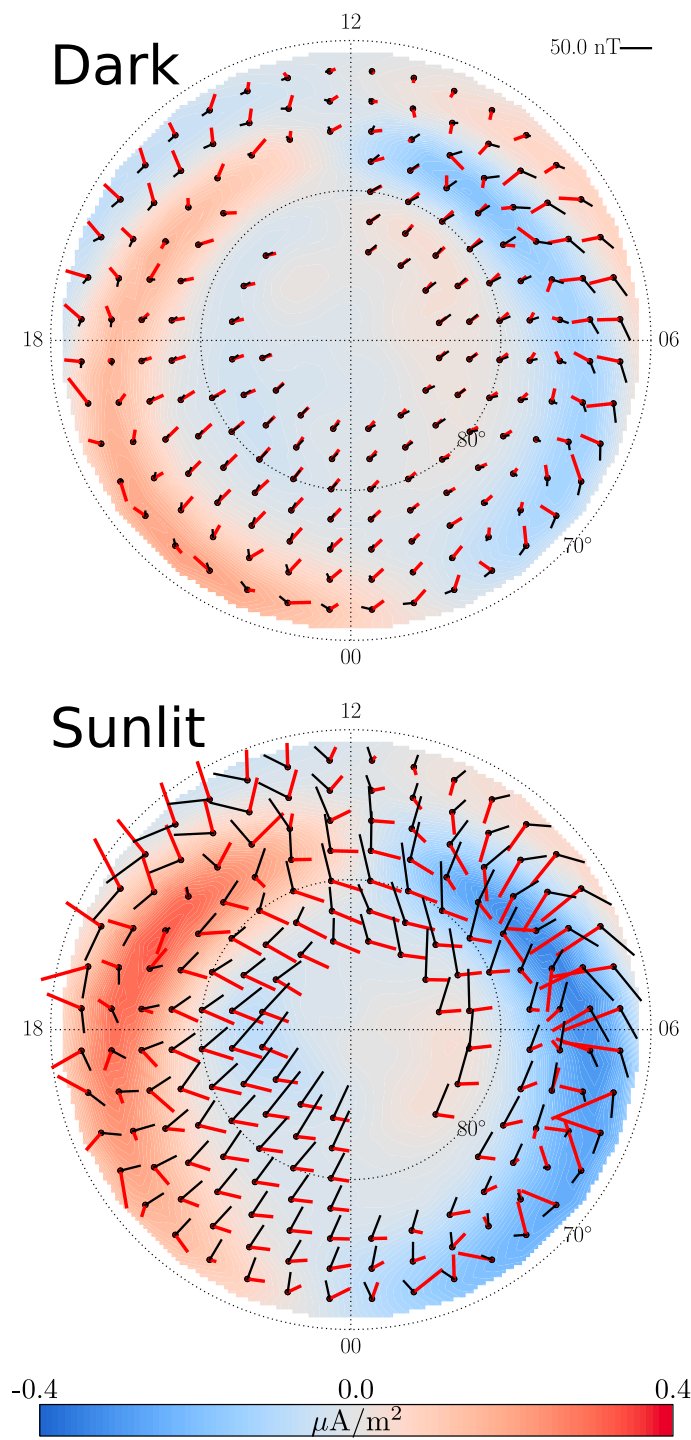

**Figure 1.** The $-\nabla_\perp \alpha$ (red vectors) calculated from equation (6), and equivalent currents $\mathbf{B}_{gnd} \times \hat{\mathbf{z}}$ (black) during (top) dark and (bottom) sunlit conditions. $-\nabla_\perp \alpha$ was calculated using an average $J_\parallel$ configuration derived from AMPERE data from periods when IMF $B_z < -1$ nT, shown in color. The equivalent currents are based on median horizontal magnetic field vectors from SuperMAG magnetometers coinciding with the AMPERE data. The medians were based on binned values in an equal area grid, the centers of which are indicated by black dots (not all cells have measurements).

## 3. Observations

In the following we calculate $-\nabla_\perp \alpha$ numerically, using equation (6) and measurements of $J_\parallel$ from AMPERE at >40° magnetic latitude. We compare the results with magnetic perturbations at high latitudes measured by SuperMAG ground magnetometers. We also compare these measurements with convection measurements from Cluster EDI.

Figure 1 shows a feather plot of $-\nabla_\perp \alpha$ (red vectors, emanating from black dots) calculated from mean Birkeland current patterns (shown in color). The AMPERE maps used to calculate the mean patterns are from the period between January 2010 and May 2013. Only data from periods when the IMF $B_z$ was less than −1 nT were used. The IMF values were obtained from the 1 min resolution OMNI data set. Figure 1 (top) is based on data from periods when the sunlight terminator (the contour of 90° solar zenith angle) was located at <70° magnetic latitude on the dayside of the noon-midnight meridian, so that the entire region poleward of 70° was in darkness. Figure 1 (bottom) is based on data from periods when the sunlight terminator was at <70° on the nightside of the noon-midnight meridian, which means that the polar region was completely sunlit. Although only the region at >70° is shown, the entire region at >40° was used in the calculations of $-\nabla_\perp \alpha$.

Also shown are median $\mathbf{B}_{gnd} \times \hat{\mathbf{z}}$ (black vectors) calculated component-wise from those SuperMAG measurements which were available at the same times as the AMPERE maps used to derive the mean Birkeland current pattern. The measurements were binned in an equal area grid, with cells centered at the positions of the black dots (not all grid cells had values). The scale of the $\mathbf{B}_{gnd} \times \hat{\mathbf{z}}$ vectors is indicated in the top right corner. We do not show the units of $-\nabla_\perp \alpha$, since they are not strictly comparable with $\mathbf{B}_{gnd} \times \hat{\mathbf{z}}$, the latter being measured in nT and $\nabla_\perp \alpha$ in Am$^{-1}$. All vectors are, however, normalized so that the top and bottom plots can be compared. We see that the currents are stronger when the ionosphere is sunlit, as expected since ionization from sunlight increases the conductivity.

The SuperMAG data were converted to quasi-dipole (QD) coordinates [*Richmond*, 1995] using the technique described by *Laundal and Gjerloev* [2014]. The AMPERE data are represented in altitude-adjusted corrected geomagnetic (AACGM) coordinates [*Baker and Wing*, 1989]. The definitions of AACGM and QD coordinates are such that at high latitudes, the coordinates are almost identical. We thus expect no significant effect of using different coordinate systems for the different data sets.

In darkness (Figure 1, top), $\mathbf{B}_{gnd} \times \hat{\mathbf{z}}$ fits $-\nabla_\perp \alpha$ remarkably well in the polar cap, in accordance with the above predictions. This is evidence that very little current flows in the polar cap and that the ground signal depends on the Birkeland currents. In the auroral zone, however, there is significant deviation between the equivalent current and $-\nabla_\perp \alpha$, probably because of enhanced conductivity from precipitation, allowing Hall currents to flow there. In sunlight the situation in the polar cap is different than for dark conditions, with a close to 90° angle between $-\nabla_\perp \alpha$ and $\mathbf{B}_{gnd} \times \hat{\mathbf{z}}$. This is also consistent with the above prediction that Hall currents dominate in this situation and Pedersen currents and $-\nabla_\perp \alpha$ cancel.

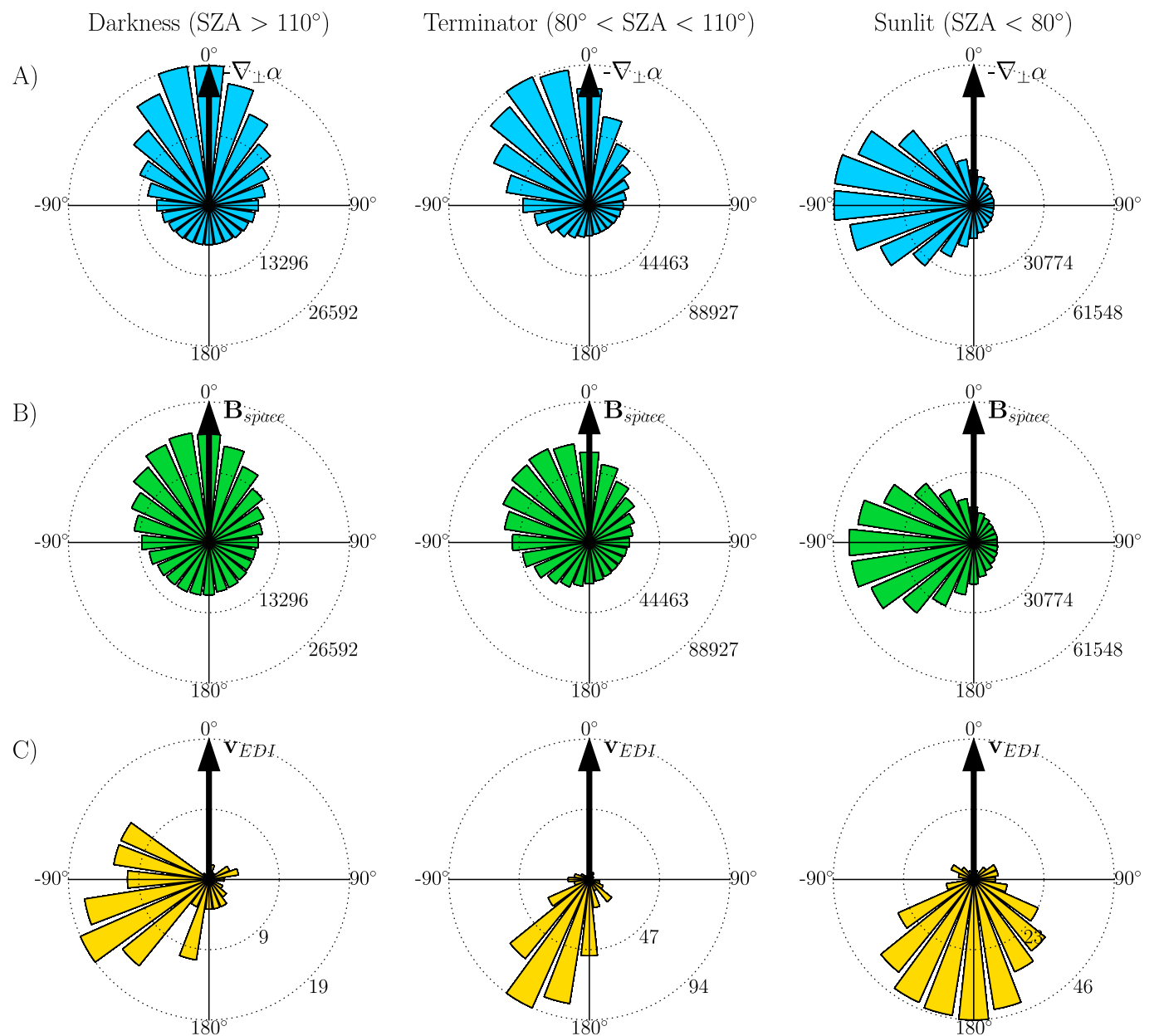

**Figure 2.** (a) Histogram of the angle of $\mathbf{B}_{gnd} \times \hat{\mathbf{z}}$ (from SuperMAG) with simultaneous $-\nabla_\perp \alpha$ directly above the ground magnetometer (calculated from AMPERE $j_\parallel$). (b) Histogram of the angle of the horizontal $\mathbf{B}_{gnd}$ (from SuperMAG) with simultaneous $\mathbf{B}_{space}$ directly above the ground magnetometer (from AMPERE). (c) Histogram of the angle of $\mathbf{B}_{gnd} \times \hat{\mathbf{z}}$ (from SuperMAG) with simultaneous $\mathbf{v}$ directly above the ground magnetometer, measured by Cluster EDI and mapped to 400 km. In all plots the vectors were measured at MLAT > 80°. The three histograms correspond to three different ranges of the solar zenith angle, indicated in the title of the plot. In Figure 2c we also required that IMF $B_z < 0$ nT.

Figure 2 shows histograms of the angle between (a) $\mathbf{B}_{gnd} \times \hat{\mathbf{z}}$ and $-\nabla_\perp \alpha$, (b) $\mathbf{B}_{gnd}$ and $\mathbf{B}_{space}$ (also from AMPERE), and (c) $\mathbf{B}_{gnd} \times \hat{\mathbf{z}}$ and the convection velocity $\mathbf{v}$, measured by Cluster EDI and mapped to 400 km and converted to AACGM coordinates [*Haaland et al.*, 2007]. The histograms are based on point by point comparisons of all ground magnetic field vectors at >80° magnetic latitude with simultaneous observations in space, within the same grid cell used in Figure 1. The count in each bin in the histogram is indicated at the concentric circles. The vector with which the ground magnetic field is compared points upward.

The histograms are sorted according to the local solar zenith angle at the magnetometer station. We see that when it is dark, the magnetic field perturbations are most commonly consistent with an equivalent current $\mathbf{J}_{eq} = \mathbf{J}_{\perp,df} \approx -\mathbf{J}_{\perp,cf} = -\nabla_\perp \alpha$ and tend to be parallel with the horizontal magnetic field perturbation in space. Figures 2a and 2b are shown with the same scale since they contain the same number of data points. We see that the distribution of angles has a larger peak and is sharper for the integrated quantity $\nabla_\perp \alpha$ compared to the distribution for the direct comparison between $\mathbf{B}_{gnd}$ and $\mathbf{B}_{space}$. For dark conditions this is consistent with $\mathbf{B}_{gnd}$ relating mostly to the global Birkeland current pattern (equation (5) with $\mathbf{J}_\perp = 0$), rather than a local overhead current. In darkness $\mathbf{B}_{gnd} \times \hat{\mathbf{z}}$ is also most often perpendicular to $\mathbf{v}$ and thus the Hall currents.

In sunlight, the equivalent currents tend to be antiparallel to $\mathbf{v}$ (i.e., parallel with Hall currents) and perpendicular to $-\nabla_\perp \alpha$ consistent with $\mathbf{J}_P \approx \nabla_\perp \alpha$. The ground magnetic field perturbations in space and on ground are typically perpendicular in sunlit conditions. This is consistent with the equivalent currents analyzed by *Friis-Christensen and Wilhjelm* [1975], which had large dawnward components at high latitude in the winter season, while the equivalent current in summer was largely sunward, antiparallel to the expected ionospheric convection. Our results also agree with the study by *Bahcivan et al.* [2013] who showed that the angle between the magnetic field perturbations and electric field vectors at Resolute Bay (approximately at 82° QD latitude) were larger in the evening than in the morning. However, the variation of the angular difference with UT in their paper was not as pronounced as the variation with local solar zenith angle shown in Figure 2c, possibly due to the seasonal variation in solar zenith angle which was not accounted for in their study.

Our interpretation of these signatures is that in darkness, there is very little current flowing in the polar cap (but $\mathbf{J}_{eq} \neq 0$), and the magnetic perturbations on ground relate mostly to the remote Birkeland currents through equation (6). In sunlight, the polar cap is conducting, and both Hall and Pedersen currents may flow there. The fact that $\mathbf{B}_{gnd} \times \hat{\mathbf{z}}$ is typically antiparallel to $\mathbf{v}$ indicates that $\mathbf{J}_H$ dominates the ground magnetic field perturbations. This is consistent with $\mathbf{J}_H \approx \mathbf{J}_{\perp,df}$ and $\mathbf{J}_P \approx \mathbf{J}_{\perp,cf}$ when the polar region is sunlit.

In the intermediate situation, when the polar cap is partially sunlit (center plots) the typical angles are somewhere in between the extremes, indicating that the ground magnetic perturbations are associated with a mix of different current systems.

## 4. Conclusions

We have shown both theoretically and with observations that in the polar cap the ground magnetic field perturbations correspond to different current systems in sunlight and in darkness. In sunlight the main contribution comes from Hall currents, which is an actual horizontal ionospheric current. In darkness, the ground magnetic perturbations are dominated by Birkeland currents in an indirect manner. The connection is indirect in the sense that Birkeland currents and horizontal curl-free currents contribute nothing in a Biot-Savart integration. However, it does affect the decomposition of the horizontal currents into curl-free and divergence-free components, and the latter can be observed.

We have focused particularly on the polar cap, since (1) the conductance there is fairly predictable, being due mostly to solar illumination, (2) the magnetic field lines are close to radial so that Fukushima's theorem applies, and (3) the flat Earth approximation used when calculating $-\nabla_\perp \alpha$ is less accurate when the region of interest is extended. We do, however, expect Birkeland current signatures in the equivalent currents also at lower latitudes. Equatorward of the auroral zone, $\sin \chi$ is still close to unity ($\sin \chi \approx 0.92$ at 50° in a dipole), and the magnetospheric driving of convection is often weak. Consequently $\mathbf{J}_\perp$ is small, so that $\mathbf{J}_{eq} \approx -\mathbf{J}_{\perp,cf}$ and the ground signal may be dominated by $-\nabla_\perp \alpha$.

As demonstrated in Figure 1, the equivalent currents have a significant dawnward component in the dark polar cap and less so in sunlight. The dawnward component has been known since before the space age [*Vestine et al.*, 1947]. Its discrepancy from the more sunward direction expected in the typical Hall currents was explained by *Vasyliunas* [1970] as being due to the difference between Hall currents and the observed divergence-free currents. The present results demonstrate the role of Birkeland currents in this decomposition. We will address the equivalent currents' dependence on sunlight in more detail in a future study.

Our results are in agreement with the actual horizontal currents in the polar cap being close to zero in darkness. In sunlight the observed vectors were found to be typically antiparallel with simultaneous measurements of high-altitude plasma convection, consistent with the Hall current system. These results imply that great care must be taken when interpreting ground magnetic field perturbations at high latitudes, such as the PC index.

**Acknowledgments**

For the ground magnetometer data we gratefully acknowledge the following: Intermagnet; USGS, Jeffrey J. Love; CARISMA, Ian Mann; CANMOS; the S-RAMP database, K. Yumoto and K. Shiokawa; the SPIDR database; AARI, Oleg Troshichev; the MACCS program, M. Engebretson; Geomagnetism Unit of the Geological Survey of Canada; GIMA; MEASURE, UCLA IGPP, and Florida Institute of Technology; SAMBA, Eftyhia Zesta; 210 Chain, K. Yumoto; SAMNET, Farideh Honary; the institutes that maintain the IMAGE magnetometer array, Eija Tanskanen; PENGUIN; AUTUMN, Martin Conners; DTU Space, Jrgen Matzka; South Pole and McMurdo Magnetometer, Louis J. Lanzarotti and Alan T. Weatherwax; ICESTAR; RAPIDMAG; PENGUIn; British Antarctic Survey; McMac, Peter Chi; BGS, Susan Macmillan; Pushkov Institute of Terrestrial Magnetism, Ionosphere and Radio Wave Propagation (IZMIRAN); GFZ, Jürgen Matzka; MFGI, B. Heilig; IGFPAS, J. Reda; University of L'Aquila, M. Vellante; SuperMAG, Jesper W. Gjerloev. We acknowledge use of NASA/GSFC's Space Physics Data Facility's OMNIWeb service and OMNI data. We thank the AMPERE team and the AMPERE Science Center for providing the Iridium-derived data products. The SuperMAG data used for this study are available at supermag.jhuapl.edu. The AMPERE data are available at ampere.jhuapl.edu. The IMF orientation data are available at omniweb.gsfc.nasa.gov. The Cluster EDI data are available at www.cosmos.esa.int/web/csa. This study was supported by the Research Council of Norway under contracts 216872/F50 and 223252/F50 (CoE).

The Editor thanks W. Jeffrey Hughes and an anonymous reviewer for their assistance in evaluating this paper.